\documentclass{cimento}

\usepackage{graphicx}  % got figures? uncomment this

\title{Dynamical properties and secondary decay effects of projectile\\
fragmentation in $^{107,124}$Sn + $^{120}$Sn collisions at 600 MeV/nucleon
}
\shorttitle{Projectile
fragmentation in $^{107,124}$Sn + $^{120}$Sn collisions 
}
\author{Jun~Su\from{ins:x},
Wolfgang Trautmann\from{ins:y}\thanks{w.trautmann@gsi.de},
Long Zhu\from{ins:x},
Wen-Jie Xie\from{ins:z}\from{ins:a},
        \atque
Feng-Shou Zhang\from{ins:b}\from{ins:c}\from{ins:d}}

\instlist{\inst{ins:x} Sino-French Institute of Nuclear Engineering and Technology,
Sun Yat-sen University, Zhuhai 519082, China
  \inst{ins:y} GSI Helmholtzzentrum f\"{u}r Schwerionenforschung GmbH, D-64291 Darmstadt, Germany 
  \inst{ins:z} Department of Physics, Yuncheng University, Yuncheng 044000, China
  \inst{ins:a} Institute of Modern Physics, Chinese Academy of Sciences, Lanzhou 730000, China
  \inst{ins:b} The Key Laboratory of Beam Technology and Material Modification of Ministry of Education,
College of Nuclear Science and Technology,
Beijing Normal University, Beijing 100875, China
  \inst{ins:c} Beijing Radiation Center, Beijing 100875, China
  \inst{ins:d} Center of Theoretical Nuclear Physics,
National Laboratory of Heavy Ion Accelerator of Lanzhou, Lanzhou 730000, China}

\begin{document}

%\PACS{24.10.Lx, 25.75.Dw}
\maketitle

\begin{abstract}

The formation of the projectile spectator and the fragmentation processes in
$^{107,124}$Sn + $^{120}$Sn collisions at 600 MeV/nucleon are studied
with the isospin-dependent quantum molecular dynamics (IQMD) model. The minimum spanning
tree algorithm and the ratio of parallel to transverse kinetic quantities are applied to identify the
equilibrated projectile spectator during the dynamical evolution. The influence of secondary decay
on fragmentation observables is investigated by performing calculations with and without the
statistical code GEMINI. The validity of the theoretical approach is examined by comparing the
calculated product yields and correlations with the experimental results of the ALADIN Collaboration for the
studied reactions. 
\end{abstract}

\section{Introduction}

The quantum-molecular-dynamics (QMD) transport model has been very successful in describing heavy-ion reactions 
at intermediate and higher energies. For 
example, the multifragmentation following collisions of $^{197}$Au + $^{197}$Au nuclei at 60 to 150 MeV/nucleon, measured by the INDRA-ALADIN 
Collaboration, has been quantitatively reproduced~\cite{zbiri07}.
The FOPI data on particle production and flow in heavy-ion reactions at energies up to 1.5 GeV/nucleon were 
successfully interpreted with the IQMD model~\cite{reis10,reis12} and information on the nuclear equation of 
state (EOS) was deduced from comparisons with IQMD~\cite{reis12,lefevre16} and UrQMD predictions~\cite{wang18}. 
The density dependence of the nuclear symmetry energy up to nearly twice the saturation density was investigated 
by analyzing elliptic flows of neutrons and light charged particles in $^{197}$Au + $^{197}$Au collisions at 400 MeV/nucleon with the UrQMD and T\"{u}bingen QMD models~\cite{russotto16,cozma18} .

A puzzle that existed for many years is posed by the apparent inability of the QMD model to describe the 
socalled rise and fall of fragment formation in projectile fragmentation at relativistic energies in the 1 GeV/nucleon 
regime with a standard minimum-spanning-tree (MST) algorithm. Begemann-Blaich {\it et al.} concluded in 1993 
that "it is not possible to reproduce the fragment distributions" in the fragmentation of $^{197}$Au nuclei 
at 600 MeV/nucleon measured by the ALADIN Collaboration~\cite{begemann93}.
A method of circumventing the problem presented by several authors is based on the early fragment recognition suggested by Dorso and 
Randrup~\cite{dorso}. Appropriate algorithms are used to identify fragmented structures that emerge in the 
colliding system at times of typically 60 fm/$c$ after the first impact. With the SACA (Simulated Annealing 
Cluster Algorithm, Ref.~\cite{puri96}) or, more recently, the FRIGA (Fragment Recognition In General Application, 
Ref.~\cite{lefevre16a}) algorithms the production of intermediate mass fragments over the full range of impact 
parameters could be satisfactorily reproduced~\cite{lefevre16a,gossiaux97,vermani09}.

In a very recent paper~\cite{sujun2018}, a study of projectile fragmentation in the 
reactions $^{107,124}$Sn + $^{120}$Sn at 600 MeV/nucleon within the framework of the IQMD model~\cite{QMD, IQMD1} 
plus GEMINI code~\cite{GEMINI1} was presented.
The IQMD-BNU (Beijing Normal University) code was used~\cite{IQMD-BNU}, a version introduced and compared to other QMD versions 
within the transport-code-comparison 
project~\cite{junxu2016}. The GEMINI model was applied to simulate the decays of the prefragments and 
the MST method was used for identifying the final products. 
Not only the properties of the projectile spectator but also the fragmentation processes 
were identified, and it was shown that the experimental results of the ALADIN Collaboration for the
studied reactions~\cite{PRC024608} are very satisfactorily reproduced.
In this contribution, a brief summary of the study is presented together with first results of an investigation 
of how to resolve the MST puzzle.

\section{The reaction model}

The theoretical framework developed for the present study is described in detail in Ref.~\cite{sujun2018}.
The parameters chosen for the IQMD description provide a compressibility of 200 MeV at saturation density
(without momentum-dependent interactions) for symmetric nuclear matter. The potential part of the symmetry 
energy has a value of 19.0 MeV at saturation density and its density dependence is modeled as a power law
with exponent $\gamma_i = 0.75$. The consequences of varying these parameters is presently under study.

The isospin and in-medium dependent parametrization of the elastic nucleon-nucleon (NN) cross sections
is taken from Refs.~\cite{cugnon96,coupland11}. To account for the fermionic nature of nucleons, 
the method of the phase-space density constraint (PSDC) of the constrained molecular dynamics (CoMD, 
Ref.~\cite{papa01}) model is applied. The phase-space occupation probability $\overline{f}_{i}$ is 
calculated by performing
the integration on a hypercube of volume $h^3$ in the phase space centered around the $i$th nucleon 
at each time step. If the phase-space occupation $\overline{f}_{i}$ has a value greater than unity, the momentum of the $i$th nucleon 
is changed randomly by many-body elastic scattering. This is done for all nucleons at each time step. In the case of collisions, the Pauli blocking of the final
states is taken into account. 
The simulations of the IQMD code are stopped when the excitation energies of the two heaviest 
prefragments are less than a specified value $E_{\rm  stop}$, chosen to have values between 2 and 4 MeV/nucleon. At that time, the GEMINI code is switched on to follow the deexcitation of the formed prefragments.

\begin{figure}
%\centerline{\includegraphics[width=11.0cm]{su-fig2} }    % Fig. 1
\centerline{\includegraphics[width=11.0cm]{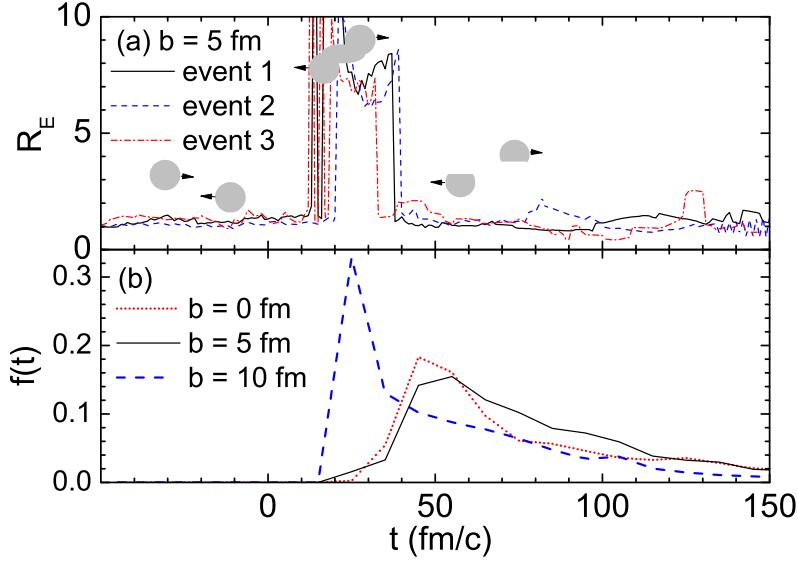} }  
\caption{(a) Temporal evolution of $R_{E}$ of the heaviest fragment (or system) in three events 
of $^{124}$Sn + $^{120}$Sn collisions at 600 MeV/nucleon with impact parameter $b = 5$~fm.
(b) Distribution of equilibration times for the projectile spectator in the same reaction 
with impact parameters $b = 0, 5,$ and 10 fm.}
\label{fig:equ_t}
\end{figure}

\section{Results and discussion}

In the IQMD model, the positions and momenta of the nucleons as a function of time can be obtained. 
At any time during the reaction process, fragments can be recognized by a minimum spanning tree (MST) 
algorithm, in which nucleons with relative distance of coordinate and momentum of 
$|r_i - r_j| \le R_0$ and $|p_i - p_j| \le P_0$ are considered as belonging to the same fragment~\cite{QMD}. The adopted values
$R_0$ = 3.5 fm and $P_0$ = 250 MeV/c are phenomenological parameters. 
The MST algorithm is performed during the collision and the mass and excitation energies of the identified
fragments are determined. The ratio of parallel to transverse quantities is further used
to distinguish the equilibrated projectile spectator:

  \begin{equation}
    R_{E}= \frac{2\sum_{i} (p_{zi} - p_{zf})^{2}}{\sum_{i} \left[(p_{xi} - p_{xf})^{2} + (p_{yi} - p_{yf})^{2} \right]},
    \label{Re}
  \end{equation}
where $p_{xi}, p_{yi}$, and $p_{zi}$ are the momentum components of the $i$th nucleon along the $x$, $y$, 
and $z$ axes; $p_{xf}, p_{yf}$, and $p_{zf}$ are the average momentum per nucleon of the fragment along the 
same axes. The summation includes all nucleons of the considered fragment.
The $z$ axis is the direction of incidence of the projectile. For an equilibrated system, the value of 
$R_E$ approaches unity. 

The temporal evolution of $R_E$ is calculated for each event in the IQMD model. As an example, the ratio
$R_E$ obtained for the heaviest fragment is shown in Fig.~1 (upper panel) for three events 
of $^{124}$Sn + $^{120}$Sn collisions at 600 MeV/nucleon with impact parameter $b$ = 5 fm.
Before the collision takes place, the identified heaviest fragment is the projectile with $R_E$ close to unity.
In the dynamical and nonequilibrium stage (from 10 to about 40 fm/$c$), the heaviest ``fragment'' is
the colliding system with a parallel energy close to the incident energy.
After the projectile spectator is separated from the participant region, it equilibrates and its value 
of $R_E$ decreases towards unity. In the actual calculations, projectile spectators with 
$0.9 < R_E < 1.2$ were regarded as equilibrated. 

The distributions of equilibration times for three impact parameters are shown 
in Fig.~1 (bottom panel). For central ($b$ = 0 fm), mid-peripheral ($b$ = 5 fm), and
peripheral ($b = 10$~fm) collisions, they are located in the interval $20 < t < 150$ fm/$c$. 
For peripheral collisions, inelastic scattering results in a peak of the  
distribution near $t$ = 25 fm/$c$. 

\begin{figure}
%\centerline{\includegraphics[width=12.0cm]{su-fig9} }    % Fig. 2
\centerline{\includegraphics[width=12.0cm]{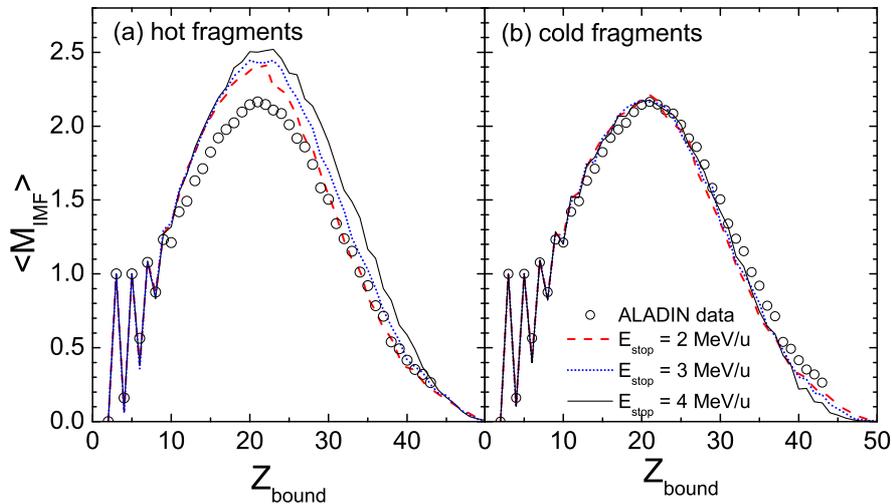} }    % Fig. 2
\caption{Mean multiplicity $\langle$$M_{\rm IMF}$$\rangle$ of intermediate-mass 
fragments ($3 \leq Z \leq 20$) as a function of $Z_{\rm bound}$ for 
reactions of $^{124}$Sn + $^{120}$Sn at 600 MeV/nucleon as obtained with (a) IQMD and (b) IQMD+GEMINI.
The calculations were performed for three values of $E_{\rm stop}$ as indicated.
The experimental data (open circles) are taken from Ref.~\protect\cite{PRC024608}.}
\label{fig:IMF}
\end{figure}

After equilibration is observed, subsequent fragmentation processes may be described statistically.
In the present case, only the final deexcitation of spectator fragments is described with
the GEMINI code. To show the role of the GEMINI code, we compare results before and after GEMINI has
been used. The correlation of the mean multiplicity of intermediate mass fragments ($3 \le Z \le 20$)
with the total bound charge $Z_{\rm bound}$ in the event is shown for $^{124}$Sn + $^{120}$Sn collisions 
at 600 MeV/nucleon in Fig. 2. The variable $Z_{\rm bound}$ represents the sum of the atomic numbers of
products with $Z \ge 2$ and, being close to the atomic number of the disintegrating projectile spectator,
is monotonically correlated with the impact parameter~\cite{sch96}.
The experimental data for the same reaction, performed with a natural Sn target (atomic weight 118.7), are
taken from Ref.~\cite{PRC024608}. The acceptance of the ALADIN forward spectrometer for projectile 
fragments is large. By studying angular distributions measured for the present reactions, 
it was found to increase with $Z$ from about 90\% for projectile fragments with $Z$ = 3 to values 
exceeding 95\% for $Z \ge 6$~\cite{sch96}.

The multiplicities exhibit the rise and fall of the fragment production. 
The lower multiplicities for smaller values of $Z_{\rm bound}$ indicate the presence of vaporization 
events in central collisions, while those in the $Z_{\rm bound} > 40$ region indicate
inelastic events in peripheral collision. The data strongly stagger for $Z_{\rm bound} < 10$. 
This phenomenon is caused by the definition of $Z_{\rm bound}$ which includes the charge of observed 
$\alpha$ particles whose number is not included in $M_{\rm  IMF}$. In particular, we observe that the 
MST method as applied in the present study is capable of reproducing the fragment production over 
the full range of $Z_{\rm bound}$, in contrast to Refs.~\cite{begemann93,gossiaux97,vermani09}. The reason for the
difference is not fully clear at present but is, most likely, 
related to the PSDC method applied to account for the Pauli principle (see below). The nearly negligible 
isotopic effect seen in a comparison of the $M_{\rm IMF}$ vs $Z_{\rm bound}$ distributions for the two 
reactions with $^{107}$Sn and $^{124}$Sn projectiles~\cite{PRC024608,sfienti2009} is also well reproduced by 
the calculations~\cite{sujun2018}.

\begin{figure}
%\centerline{\includegraphics[width=13.0cm]{su-fig12} }    %  Fig. 3
\centerline{\includegraphics[width=13.0cm]{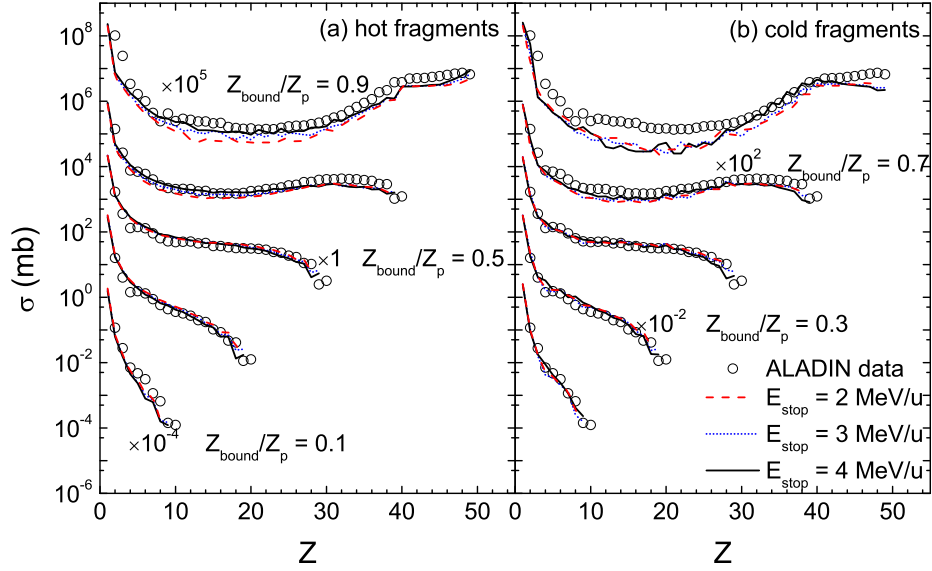} }    %  Fig. 3
\caption{Calculated cross sections d$\sigma$/d$Z$ for the fragments produced in the reaction 
$^{124}$Sn + $^{120}$Sn at 600 MeV/nucleon in comparison with the experimental data.
The events are sorted into five intervals of $Z_{\rm bound}/Z_{p}$ with centers as indicated and width 0.2 
($Z_{p} = 50$ is the atomic number of the projectile).
The calculations with IQMD and with IQMD+GEMINI are shown in the left and right panels, respectively. The
experimental data are taken from Ref.~\protect\cite{PRC024608}.}
\label{fig:dMdz}
\end{figure}

The cross sections d$\sigma$/d$Z$ obtained for the fragments produced in the $^{124}$Sn + $^{120}$Sn reaction 
are shown in Fig. 3 after sorting into five intervals of $Z_{\rm bound}$. 
The calculations are again shown with and without the use of GEMINI.
Overall, the model calculations reproduce the data rather satisfactorily with a quality comparable to that of 
the statistical description presented by Ogul {\it et al.}~\cite{PRC024608} or Mallik {\it et al.}~\cite{mallik2011}. 
Only the experimental cross sections of the most peripheral collisions ($Z_{\rm bound} \ge 40$) are somewhat underrepresented. 
The effect of choosing different values of $E_{\rm stop}$ is  very small, as illustrated
in Figs. 2 and 3. It may be concluded that the deexcitation of moderately excited 
fragments with excitation energies between 2 and 4 MeV/nucleon is adequately reproduced with both, 
the IQMD and GEMINI codes. The main effect of GEMINI, as observed in the figures, is the small
reduction of the fragment multiplicity, caused by the breakup of some of the lighter fragments
in the final deexcitation stage.

\begin{figure}
%\centerline{\includegraphics[width=11.0cm]{eff_PSDC} }    %  Fig. 4
\centerline{\includegraphics[width=11.0cm]{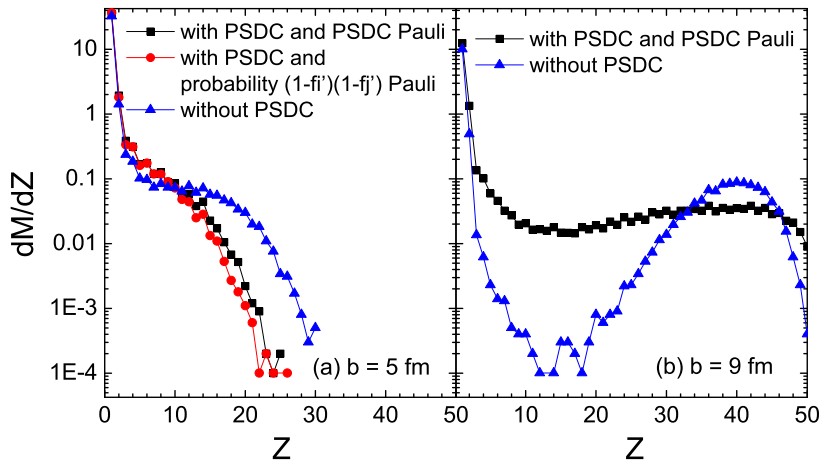} }    %  Fig. 4
\caption{Differential fragment multiplicity d$M$/d$Z$, calculated with the indicated methods to account 
for the Pauli principle, in the reaction 
$^{124}$Sn + $^{120}$Sn at 600 MeV/nucleon for impact parameters $b = 5$~fm (a) and $b = 9$~fm (b).
}
\end{figure}

\section{PSDC and fragmentation}

In the case of collisions, the Pauli blocking of the final states has to be taken into account. Different algorithms are 
being used for this purpose.
In the original IQMD~\cite{IQMD1}, collisions are allowed with the probability (1-$f'_{i}$)(1-$f'_{j}$); here 
$f'_{i}$ and $f'_{j}$ are the phase space densities at the final states before the scattered particle is placed there.
The Pauli blocking method related to the PSDC is used in the present work.
These two methods have been compared in our previous work \cite{PRC017602}.
The Pauli blocking method related to the PSDC is consistent with our using the PSDC method throughout the reaction
to account for the fermionic nature of nucleons.
It is required that no state with $\overline{f}_{i}$ $>$ 1 is created in a binary NN collision, otherwise it will be 
changed by the PSDC method.
Thus, for each binary NN collision the phase space occupation probabilities $\overline{f}_{i}$ and $\overline{f}_{j}$ at the final states are 
measured.
Only if $\overline{f}_{i}$ and $\overline{f}_{j}$ at the final states are both less than 1, the scattering is accepted.

The effects on fragment production following from choosing either one or the other method are illustrated in Fig. 4 
with calculations performed for fixed impact parameters $b = 5$ and 9~fm. In particular at the larger impact parameter, 
the obtained multiplicities of intermediate-mass fragments depend strongly on this choice. They are systematically lower 
in calculations without PSDC than with PSDC, while the yield of heavy residue-type fragments is larger 
if the PSDC is not applied (blue triangles). Whether the probability method or the PSDC is used for collisions seems 
to have a minor effect (red dots and black squares, respectively, in the left panel). The underlying reason for the 
strong consequences following from applying the PSDC at each time step is still under investigation. 

\acknowledgments
This work was supported by the National Natural Science Foundation of China under Grants No. 11405278, No. 11605296, 
No. 11505150, and No. 11635003, the Natural Science Foundation of Guangdong Province China under Grant No. 2016A030310208, 
and the China Postdoctoral Science Foundation under Grant No 2015M582730.
The authors are grateful to the ALADIN Collaboration for providing the numerical values of experimental results 
reported in Ref.~\cite{PRC024608}.

\end{document}